\documentclass[preprint,review,12pt]{elsarticle}
\biboptions{authoryear}

\usepackage{graphicx}
\usepackage{amsmath}
\usepackage{amssymb}
\usepackage{lipsum}
\usepackage{comment}
\usepackage{xurl}
\usepackage{float}
\usepackage{tabularray}
\usepackage{threeparttable} 
\usepackage{adjustbox}
\usepackage{tabularx}
\usepackage{xcolor}
\usepackage[normalem]{ulem}

\usepackage{lineno}

\journal{Geochimica et Cosmochimica Acta}

\begin{document}

\begin{frontmatter}



\title{First experimental determination of the  $^{40}$Ar($n,2n$)$^{39}$Ar reaction cross section and  $^{39}$Ar production in Earth's atmosphere}


\author[first]{S. Bhattacharya}
\affiliation[first]{The Hebrew University of Jerusalem,
            addressline={}, 
            city={Jerusalem}, 
            postcode={ 91904},
            country={Israel},}
\author[first]{M. Paul\corref{cor1}\fnref{label2}}
\ead{paul@vms.huji.ac.il}
\author[first]{R. N. Sahoo} 
\author[second]{R. Purtschert}
\affiliation[second]{University of Bern,
            addressline={},
            postcode={ 3012},
            city={Bern}, 
            country={Switzerland},
            }
\affiliation[third]{Technical University Dresden,
            addressline={},
            city={Dresden},
            country={Germany},
            }
\author[third]{H.F.R. Hoffmann} 
\author[third]{M. Pichotta} 
\author[third]{K. Zuber} 
\affiliation[fourth]{Helmholtz-Zentrum-Dresden-Rossendorf,
            addressline={},
            city={01328 Dresden},
            country={Germany},
            }
\author[fourth]{D. Bemmerer} 
\author[fourth]{T. D\"oring} 
\author[fourth]{R. Schwengner}            
\author[fifth]{M.L. Avila}
\affiliation[fifth]{Argonne National Laboratory,
            addressline={},
            city={Argonne},
            state={IL},
            postcode={ 60439},
            country={USA},
            }
\author[fifth]{E. Lopez-Saavedra}
\author[fifth]{J.C. Dickerson}
\author[fifth]{C. Foug\`eres}
\author[fifth]{J. McLain}
\author[fifth]{R.C. Pardo}
\author[fifth]{K.E. Rehm}
\author[fifth]{R. Scott}
\author[fifth]{I. Tolstukhin}
\author[fifth]{R. Vondrasek}
\affiliation[sixth]{University of Notre Dame,
            addressline={},
            city={Notre Dame},
            state={IN},
            postcode={ 46556},
            country={USA},
            }
\author[sixth]{T. Bailey}
\author[sixth]{L. Callahan}
\author[sixth]{A.M. Clark}
\author[sixth]{P. Collon}
\author[sixth]{Y. Kashiv}
\author[sixth]{A. Nelson}
\author[sixth]{D. Robertson}
\affiliation[seventh]{University of Illinois 
Chicago, 
            addressline={},
            city={Chicago},
            state={IL},
            postcode={ 60607}, 
            country={USA}, 
            }          
\author[seventh]{D. Neto}
\author[seventh]{C. Ugalde}
\affiliation[eighth]{Soreq Nuclear Research Center,
            addressline={},
            city={Yavne},
            postcode={ 81800},            
            country={Israel},
            }
\author[eighth]{M. Tessler}
\author[eighth]{S. Vaintraub}

\begin{abstract}
The cosmogenic $^{39}$Ar(t$_{1/2}$= 268 years)  isotope of argon is used for geophysical dating and tracing of underground and ocean water, as well as ice owing to its appropriate half-life and chemical inertness as a noble gas; $^{39}$Ar serves also in nuclear weapon test monitoring. We measured for the first time the total cross section of the main $^{39}$Ar cosmogenic production reaction in the atmosphere, namely 
$^{40}$Ar$(n,2n)^{39}$Ar, using 14.8$\pm0.3$ MeV neutrons. The neutrons, produced by a deuterium-tritium generator, impinged on a stainless steel sphere filled with Ar gas highly enriched in the $^{40}$Ar isotope and were monitored by a stack of fast-neutron activation foils. The reaction yield was measured by atom counting of long-lived $^{39}$Ar with noble gas accelerator mass spectrometry and, independently, by decay counting relative to atmospheric argon ($^{39}$Ar/Ar= $8.12\times 10^{-16}$). A total $^{40}$Ar$(n,2n)^{39}$Ar cross section of 610$\pm100$ mb was determined at 14.8$\pm0.3$ MeV incident neutron energy. This result serves as a benchmark for recent  
theoretical calculations and evaluations, found to reproduce well the experimental total cross section. We use these energy-dependent theoretical cross sections together with experimental spectra of cosmogenic neutrons at different altitudes to calculate the global average rate of neutron-induced $^{39}$Ar atmospheric production, resulting in $770\pm240$ $^{39}$Ar  atoms/cm$^2$/day. The secular equilibrium between the $^{39}$Ar calculated production rate and radioactive decay rate leads to a partial isotopic abundance $^{39}$Ar/Ar$= (5.9\pm 1.8) \times 10^{-16}$, showing that $\approx$73\% of atmospheric $^{39}$Ar is produced by cosmogenic neutrons, the remaining part believed to be induced by muons and high-energy $\gamma$ rays. The $^{40}$Ar($n,2n$)$^{39}$Ar cross section at 14 MeV is also a key parameter for quantifying the anthropogenic contribution to atmospheric $^{39}$Ar produced during the thermonuclear tests of the 1960s. We estimate that anthropogenic $^{39}$Ar accounts for roughly 20\% of the present atmospheric inventory.
\end{abstract}




\end{frontmatter}



\begin{linenumbers}
\section{Introduction}
\label{introduction}
The radioactive $^{39}$Ar nuclide with a half-life of 268 years \citep{STO65,GOL23} occurs in nature owing to continuous production in the atmosphere by cosmic ray bombardment and in the lithosphere by both cosmogenic and radiogenic processes. 
It is also artificially produced in nuclear explosion tests. Owing to its half-life and chemical inertness, the radioactive $^{39}$Ar nuclide is widely used in geophysical (hydrological dating) and environmental (nuclear forensics) research. Saldanha \textit{et al.}
\citep{SAL19} have reviewed the sea-level $^{39}$Ar atmospheric production modes among which the dominant reaction is $^{40}$Ar$(n,2n)^{39}$Ar, together with a weaker $^{40}$Ar$(n,pn)^{39}$Cl($t_{1/2}$= 56 min) component which feeds $^{39}$Ar via $\beta$ decay. The $^{40}$Ar$(n,2n)^{39}$Ar reaction, with a neutron energy threshold of 10.12 MeV, occurs mainly in the upper atmosphere from the interaction of fast secondary cosmic neutrons with the stable isotope  $^{40}$Ar of argon (atmospheric $^{40}$Ar isotopic abundance of 99.6\%). Lithospheric production  involves reactions induced by atmospheric and radiogenic neutrons as well as muogenic processes (see \citep{MUS23} for a recent review). 
\begin{table}[H]
\centering
{\small %
\begin{threeparttable}
\caption{\label{cross-section} Theoretical (theo), evaluated (eval) and experimental (exp) values of the $^{40}$Ar(n,2n)$^{39}$Ar reaction cross section ($\sigma$). Theoretical and evaluated values are calculated at 15 MeV. Experimental values are measured at the specified energy. Uncertainties of experimental values (in parenthesis) correspond to one sigma. Note that the two experimental values of \citep{MAC12} correspond to two different neutron energies given in the footnotes.}
\begin{tabular}{lccc}
\hline
\hline
 & theo/eval/exp &  $\sigma$ (mb) & ref 
\\
\hline\\
ACTIVIA\tnote{a} \ (2008) &theo & 10.1 & \citep{BAC08}\\
INCL++(ABLA 07) (2014)& theo & 438\tnote{b}  & \citep{MAN14} \\
TALYS-2.00 (2023) & theo & 649 & \citep{KON23} \\
ENDF/B-VIII.1 (2024) & eval & 667 & \citep{NOB24} \\
ENDF/B-VII.1 (2011)& eval & 900 & \citep{CHA11} \\
JENDL-5 (2021) & eval & 900 & \citep{IWA23} \\
MacMullin \textit{et. al.} (2012)\tnote{c} & exp & 80(20)\tnote{d} & \citep{MAC12} \\
\hspace{2cm}"  & exp  & 130(20)\tnote{e} & \citep{MAC12} \\
  This work\tnote{f} & exp & 610 (100) \\
\hline
\end{tabular} 
\begin{tablenotes}
    \item[a] Based on \citep{SIL73I,SIL73II}
    \item[b] Extracted from \citep{SAL19}
    \item[c] Partial cross section to E$_x$($^{39}$Ar)= 1267 keV
    \item[d] E$_n=$13.4(8) MeV
    \item[e] E$_n=$15.0(9) MeV
    \item[f] Total cross section for neutron energy 14.8(3) MeV
\end{tablenotes}

\end{threeparttable}
}
\end{table} 

The total cross section of the $^{40}$Ar$(n,2n)^{39}$Ar fast-neutron reaction has  not been heretofore measured and is the object of this article in which it is determined experimentally at a neutron energy of $14.8\pm0.3$ MeV. Partial cross sections, populating two low-lying  $^{39}$Ar$^*$ excited states were measured in  \citep{MAC12} by neutron activation and $\gamma$-spectrometry. Saldanha et al. \citep{SAL19} measured the  $^{40}$Ar$(n,2n)^{39}$Ar production yield for a neutron spectrum mimicking that of sea-level cosmogenic neutrons. Theoretical cross-section values from different models at 15 MeV are listed in Table \ref{cross-section}, together with the experimental results from \citep{SAL19}  and from this work. 
At higher altitudes the neutron spectra are shifted towards higher neutron energies for which cross section data have so far been largely unavailable. They can be inferred from recent theoretical models, based on the fair agreement with our experimental value at 15 MeV.  Validating the theoretical cross-section calculations will allow also for a more precise reconstruction of the historical $^{39}$Ar input function \citep{CEN95shrt,GU21, LOO68}, which is crucial for using this isotope in different dating applications in the hydro and cryosphere \citep{COR07,YOK12}. Dating of air trapped in shallow ice , for example in blue-ice regions \citep{BUI14} or high altitude glaciers \citep{HOU25}, could be biased by cosmogenic $^{39}$Ar \textit{in-situ}  production on various target elements, including Ar trapped in air bubbles. Our experimental value of the $^{40}$Ar$(n,2n)^{39}$Ar reaction cross section plays also an essential role in the quantitative estimate of the anthropogenic $^{39}$Ar input to the environment during the thermonuclear atmospheric tests releasing 14.1 MeV neutrons from deuterium-tritium fusion \citep{LOO68,GU21}. Experimental knowledge of the relevant $(n,2n)$ reaction will allow a more reliable assessment of this contribution to the $^{39}$Ar atmospheric inventory. 

An independent incentive for the present measurement came from an experiment using laser-induced inertial confinement fusion performed at National Ignition Facility (see a review in \citep{CER18}). In this experiment Ar seeds were added to a deuterium-tritium (DT) capsule to study neutron-induced reactions in a high-density and high neutron density plasma produced during implosion. The $^{40}$Ar$(n,2n)^{39}$Ar reaction serves there as an internal monitor of the 14 MeV neutrons produced by DT fusion.\\  

Section \ref{Experimental} describes our experimental method based on 14.8 MeV neutron activation of $^{40}$Ar and two independent methods of detection of $^{39}$Ar, namely atom counting by noble gas accelerator mass spectrometry (NOGAMS) and low-level counting (LLC) of its decay $\beta$'s. Section \ref{results} presents our cross section results and these results are discussed in Section \ref{discussion} in the context of $^{39}$Ar cosmogenic atmospheric production.

\section{Materials and Methods}
\label{Experimental}
\subsection{Sample preparation and  neutron activation}
\label{Sample}
\begin{figure}[H]
	\centering 
    \includegraphics[width=0.7\textwidth, angle=0]{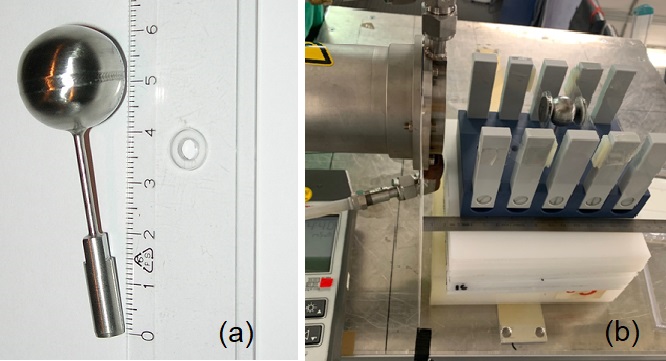}	
    \caption{Photograph of (a) $^{40}$Ar-filled sphere used for activation at the TUD DT neutron generator; (b) activation setup (from left to right): accelerator beam pipe ending with the beam stop holding the tritium target, two rows of (light grey) rectangular neutron absorbers, the 20 mm diameter sphere sandwiched between Al, Zr and Nb monitor foils of same diameter. The sphere is positioned at 10 cm from the beam stop holding the tritium target.
}
    \label{HZDR_act}
\end{figure}

 Argon samples enriched in $^{40}$Ar were prepared for activation at Helmholtz Zentrum Dresden-Rossendorf (HZDR) by filling cryogenically two spheres, denoted as HZDR and HZDR23. The spheres (Fig. \ref{HZDR_act}a), made of stainless steel, are 20 mm in inner diameter with 0.6 mm wall thickness and
sealed by a 3 mm diameter spring loaded ball that can be released by a set screw to transfer the gas to a secondary container after irradiation. Table \ref{irradiation} lists the properties of the samples.  
The HZDR and HZDR23 spheres were separately activated in 2021 and 2023, respectively, by neutrons from the D-T neutron generator \citep{KLI18} of Technical University Dresden  (TUD). The neutrons were generated by a 300 keV deuteron beam bombarding a planar tritium thick target of titanium hydride with a diameter of 1 cm. The spheres (Fig. \ref{HZDR_act}b), sandwiched between activation monitors appropriate for fast neutrons (Nb, Zr and Al foils of 20 mm diameter) were positioned as Nb-Zr-Al-sphere-Al-Zr-Nb in the forward direction 10 cm from the accelerator beam stop holding the tritium target. The Zr monitors, used to determine the effective energy of the neutron beam via the energy-dependent $^{90}$Zr$(n,2n)^{89}$Zr cross section \citep{NOB24}, yielded values of 14.67$\pm$0.15 MeV and 15.0$\pm$0.4 MeV for the 2021 and 2023 campaigns, respectively.
We adopt the unweighted average 14.8 $\pm$ 0.3 MeV of the experimental values  as representative of both experiments. The value is also consistent with that  of the mean and standard deviation calculated from the experimental excitation function of the $t(d,n)^4$He reaction \citep{DAV57GCA} between the deuteron bombarding energy of 300 keV and zero for the thick target used in our experiment. The irradiation
duration per campaign was 5 h (including an interruption of 1 h for the HZDR23 sample). The neutron field intensity varied over time by
$\approx30$\% as measured by the deuteron beam intensity on target. The time profile and decay
corrections were later taken into account in the analysis. After each activation, the activity of the 
monitor samples was measured using a lead-shielded germanium detector with a distance of
5 cm (10 cm) between crystal and sample in the 2021 (2023) experiments. The germanium detector’s full
energy efficiency was calibrated using point-shaped reference sources. Correction factors
for geometry and $\gamma$ self absorption in the monitor foils were obtained from  detailed GEANT4 simulations \citep{AGO03shrt}.
The integral neutron fluences (see Table \ref{irradiation}) were determined by the Nb (0.7242 g/cm$^2$) and Al (0.2303 g/cm$^2$) monitors. The cross section of the  
$^{93}$Nb$(n,2n)^{92m}$Nb(10.15 d) reaction (0.466$\pm$0.019) b \citep{KOS23} and of the 
$^{27}$Al$(n,\alpha)^{24}$Na(14.96 hr) reaction (0.1216$\pm$0.0051) b \citep{MAN07} were used. These experimental cross sections and their uncertainties represent well the spread of
available evaluated cross sections at our effective neutron energy. 
For each of the Nb and Al sample pairs, the geometric mean
of both fluence values describes the fluence in their center assuming exponential neutron
attenuation. The fluences extracted from the Al
and Nb monitors 
are consistent and were combined as a weighted mean and its variance (Table \ref{irradiation}). The one-sigma systematic uncertainty (2.0\%) of the full energy efficiency is determined conservatively
as the maximum uncertainty of the calibration sources’ absolute activity. Uncertainties in monitor mass ($<1$\%), literature half-lives contribution (0.2\%) and counting statistics (0.2\%) are taken into account.
\begin{table}[H]
\centering
\begin{threeparttable}
{\small %
\caption{\label{irradiation} Activated Ar samples. Uncertainties in fluence values correspond
to one sigma.}}
\begin{tabular}{lccc}
\hline
\hline
 & SNRC & HZDR & HZDR23\\
 & quartz & stainless &  stainless\\
 & ampoule& steel sphere & steel sphere\\
\hline\\
$^{40}$Ar ($\%$) & 0.006 & 99.992 & 99.992\\
$^{38}$Ar ($\%$) & 99.961 & 0.004 & 0.004\\
$^{36}$Ar ($\%$) & 0.033 & 0.004 & 0.004\\
Ar mass (g) & 0.0053 & 0.3903 & 1.0176 \\
neutron fluence (cm$^{-2}$) & $\approx2\times 10^{14}$ $^{a}$ & $6.92(24) \times 10^{11}$ $^{b}$ & $7.74(56)\times 10^{11}$ $^{b}$ \\
\hline
\end{tabular}
\begin{tablenotes}
        \item $^{a}$ Thermal neutron activation at SNRC. Sample SNRC was diluted after activation with $^{38}$Ar(99.96\%) and $^{nat}$Ar to a final isotopic ratio of $^{38}$Ar$/^{40}$Ar$\approx9.5$\% for the NOGAMS setup and $^{38,40}$Ar transmission measurements; see \ref{NOGAMS}.
        \item $^{b}$ 14.8$\pm0.3$ MeV neutron activation at the TUD neutron generator.
    \end{tablenotes}
\end{threeparttable}
\end{table}

After appropriate radioactive cooling, each sphere was shipped to Argonne National Laboratory (ANL) and its argon content was separated in two aliquots, one to be analyzed for its $^{39}$Ar/$^{40}$Ar isotopic ratio by NOGAMS at ANL and the second for its $\beta$ activity by LLC at University of Bern.\\
In addition, an Ar sample highly enriched in $^{38}$Ar contained in a small quartz ampoule (Table \ref{irradiation}) was activated at the Soreq Nuclear Research Center (SNRC) with thermal neutrons from a 5 MW reactor to produce $^{39}$Ar by $^{38}$Ar$(n,\gamma)^{39}$Ar. After activation, the SNRC sample was diluted with $^{38}$Ar and $^{nat}$Ar   to obtain a sample volume and isotopic abundance appropriate for establishing the  NOGAMS and LLC $^{39}$Ar detection conditions  and to measure the accelerator transmission efficiencies for $^{38}$Ar and $^{40}$Ar (see Section \ref{NOGAMS}).

\subsection{$^{39}$Ar detection}
\label{detection}
\subsubsection{NOGAMS analysis}
\label{NOGAMS}
\begin{figure}[H]
	\centering 
    \includegraphics[width=0.65\textwidth, angle=0]{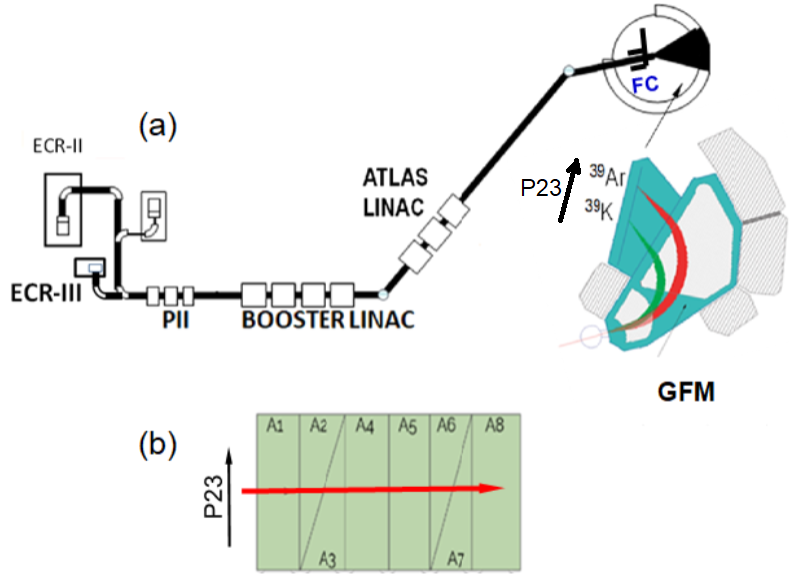}	
\caption{Schematic diagram of: (a) the Noble Gas Accelerator Mass Spectrometry setup at Argonne National Laboratory for $^{39}$Ar detection. $^{38,39,40}$Ar$^{8+}$ ions are sequentially injected from the Electron Cyclotron Resonance (ECR-III) ion source and accelerated through the Positive Ion Injector (PII) and Booster- and ATLAS-Linac. Intensity of stable isotope $^{38,40}$Ar$^{8+}$ accelerated beams are measured as charge current (typically of the order of nA) in an electron-suppressed Faraday Cup (FC). The rare isotope $^{39}$Ar is counted as individual ions in the Enge Gas-Filled  Magnetic  spectrograph (GFM); (b) the multi-anode plate of the Monica ionization chamber \citep{CAL22shrt} located perpendicular to the focal plane of the GFM. Energy loss signals for ions entering the detector are extracted from each anode A1-A8 for identification. The position P23 along the GFM dispersion axis is measured by the normalized difference of signals in the split anodes A2 and A3 (P23= (A3-A2)/A3+A2)).}
	\label{NOGAMS_MONICA}
\end{figure}

Noble gas accelerator mass spectrometry, an offshoot of standard accelerator mass spectrometry (AMS) \citep{KUT23}, is an atom counting technique based on positive ion injection,  enabling ultra-sensitive detection of noble gas isotopes; the latter do not form stable negative ions used in standard AMS and are excluded thereof. The technique was developed at Argonne National Laboratory (ANL) for the detection of $^{39}$Ar and geophysical 
dating  \citep{COL04a} and has been applied since to nuclear astrophysics \citep{TES18shrt,NAS05shrt}; see \citep{PAU19} for a detailed description. Fig. \ref{NOGAMS_MONICA}(a) shows a 
schematic illustration of the facility at the ATLAS accelerator at ANL. The activated gas samples (Table \ref{irradiation}) are loaded into the ECR3 Electron Cyclotron Resonance ion source and highly charged Ar$^{8+}$ ions are extracted. Following mass-to-charge ($m/q)$ magnetic analysis, $^{38,39,40}$Ar ions are sequentially injected and accelerated by the ATLAS superconducting linear accelerator, which acts as an additional $m/q$ filter, to an energy of 5.3 MeV/nucleon. The stable $^{38,40}$Ar$^{8+}$ charge current (of the order of nA) is measured in an electron-suppressed Faraday cup; quantitative attenuation of the more intense $^{40}$Ar$^{8+}$ current is applied (see Table \ref{table-AMS}). $^{39}$Ar and parasitic ions (isobaric ions and ions with close-by $m/q$ ratio transported by the accelerator), with rates of 1--100 s$^{-1}$ in our experiments,  are analyzed by the gas-filled Enge spectrograph \citep{PAU89}. The ions are  spatially separated therein as  function  of their ratio of
 mass to mean ionic charge ($m/\bar{q}$); the latter results from atomic collisions which change the ionic charge in the gas-filled magnet. Owing to their energy of a few MeV/nucleon, the ions are further identified by energy loss measurements in the Monica  position-sensitive multi-anode ionization chamber \citep{CAL22shrt}. Dispersion (spatial separation) along the focal plane is measured by the parameter $P23= (A3-A2)/(A3+A2)$ where Ai denotes here the energy loss signal in the corresponding anode (Fig. \ref{NOGAMS_MONICA}(b)).
\begin{figure}[H]
	\centering 
	\includegraphics[width=0.8\textwidth, angle=0]{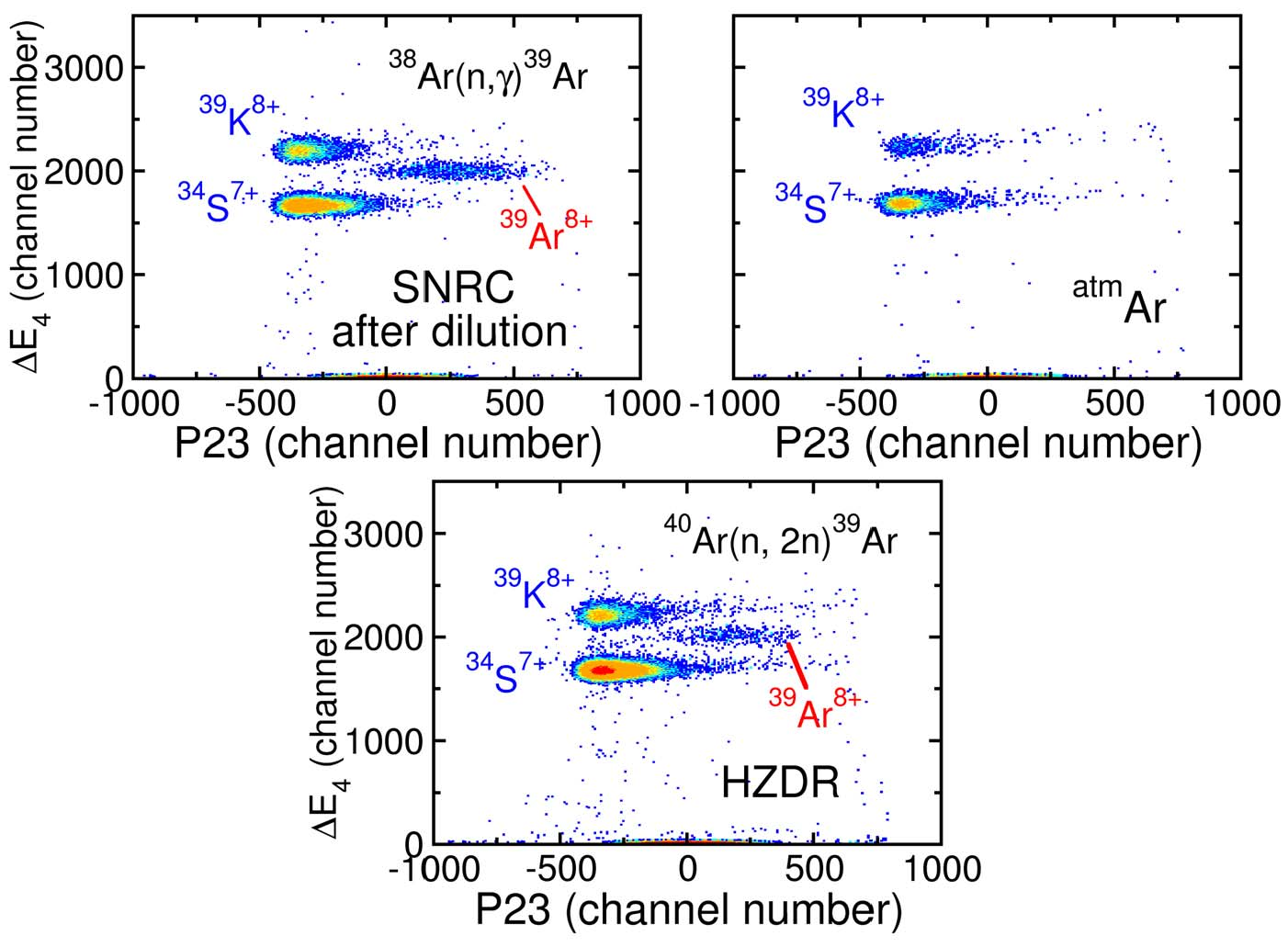}	
	\caption{Identification spectrum of $^{39}$Ar ions in the detector
measured for: (top left) the diluted SNRC sample irradiated at Soreq NRC; (top right) a non-irradiated atmospheric Ar sample ($^{39}$Ar/Ar= $8.12\times 10^{-16}$) where $^{39}$Ar is below detection sensitivity; (bottom) the HZDR (2022) irradiated sample. The horizontal axis represents the dispersion along the focal plane and the vertical axis represents the differential energy loss signal measured in the fourth anode of the focal-plane ionization chamber. The upper and lower left-hand groups originate from $^{39}$K$^{8+}$  (stable isobar of $^{39}$Ar) and $^{34}$S$^{7+}$ ions, respectively, both chemical impurities in the ion source. $^{34}$S$^{7+}$ ions are nearly degenerate in $m/q$ with $^{39}$Ar$^{8+}$.} 
	\label{HZDR_AMS}
\end{figure}

Fig. \ref{HZDR_AMS} displays an example of identification spectra where the group of $^{39}$Ar ions, clearly separated from isobaric $^{39}$K ions and parasitic $^{34}$S ions (originating from chemical impurity in the ion source), can be counted. A similar spectrum accumulated when a $^{nat}$Ar sample was loaded in the ion source shows a negligible background in the region of $^{39}$Ar interest, consistent with the atmospheric $^{39}$Ar cosmogenic abundance ($^{39}$Ar/Ar= $8.12\times 10^{-16}$ \citep{LOO83,BEN07shrt,GOL23}) below the sensitivity of the present measurements.
\begin{table}[H]
{\small %
\centering
\caption{\label{table-AMS}$^{39}$Ar/Ar isotopic abundance and $^{40}$Ar$(n,2n)$$^{39}$Ar extracted cross section value for irradiated HZDR and HZDR23 samples in NOGAMS and LLC measurements. Final values were rounded to relevant number of digits. Uncertainties (in parenthesis) correspond to one sigma.}}\

\begin{tabular}{l c c c }
\hline
\hline
 & \multicolumn{2}{c}{HZDR} & HZDR23 \\
 & 2021 & 2022 & 2023 \\
\hline
NOGAMS & & &\\
\footnotesize{Ave. FC $^{40}$Ar$^{8+}$ (nA)} & 1.8(1) & 1.75(3) & 2.6(1)\\
\footnotesize{$^{40}$Ar$^{8+}$ attenuation factor} & 26.7(8) & 26.8(8) & 28.0(8) \\
\footnotesize{Ave. $^{39}$Ar rate-gross (cpm)}   &0.96(5) & 0.92(4) &1.56(7) \\
\footnotesize{Ave. $^{39}$Ar rate-net (cpm)}   &0.70(4) & 0.89(4) &1.50(7) \\
\footnotesize{$^{39}$Ar fractionation correction}& 1.13(10) & 1.08(11) & 1.03(5) \\
\footnotesize{R$_{net}^a$ ($10^{-13}$) -  NOGAMS}  & 3.5(3) & 4.4(5) & 4.6(4) \\
\footnotesize{cross section (mb)- NOGAMS} & 510(50) & 640(80) &  590(70) \\
\hline
LLC & & & \\
\footnotesize{sample  count rate-gross (cpm)} &  & 0.455(5) & 0.532(9)\\
\footnotesize{sample count rate-net (cpm)} &  & 0.319(6) & 0.409(9)\\ 
\footnotesize{atmosphere-net (cpm)} &  & 0.059(9) & 0.058(6)\\ 
\footnotesize{R$_{net}^a$(sample)/R$_{net}^a$(atmosphere)} &  & 460(70) & 727(60)\\ 
\footnotesize{R$_{net}^a$(count.  gas)/R$_{net}^a$(atmosphere)}&  & 5.5(8) & 7.1(7)\\ 
\footnotesize{R$_{net}^a$ ($10^{-13}$) -
 LLC} &  & 3.7(6) & 5.9(4) \\
\footnotesize{cross section (mb) -  LLC} & & 540(90) & 760(80) \\
\hline
grand average cross section (mb) &\multicolumn{3}{c}{ 610(100)}  \\
\hline
\end{tabular}
\begin{tablenotes}
        \item \footnotesize{$^a$} background subtracted $^{39}$Ar/Ar isotopic abundance
   \end{tablenotes}
\end{table}

The (absolute) $^{39}$Ar/$^{40}$Ar isotopic abundance measured by NOGAMS in each of the experiments (2021, 2022, 2023) listed in Table \ref{table-AMS} 
is determined by the ratio of the $^{39}$Ar count rate (s$^{-1}$) measured in the Monica detector and the $^{40}$Ar$^{8+}$ ion rate  $i$/8$e$, where $i$ is the charge current in A measured in the Faraday cup and $e$ is the electron charge in C, respectively. The charge current $i$ is corrected for the beam attenuation factor. An additional fractionation correction (see Table \ref{table-AMS}) is applied to take into account the ratio of $^{39}$Ar and $^{40}$Ar accelerator transmission efficiencies. The $^{39}$Ar transmission efficiency is determined in each experiment by the average transmission efficiency for $^{38}$Ar and $^{40}$Ar using the diluted SNRC sample; the latter (of the order of 50\%) are themselves measured as ratio of charge currents after the ion source and at the magnetic spectrograph entrance. The ion transmission efficiency between the Faraday cup and the Monica detector was established to be 100\%. 
\subsubsection{Low-level counting analysis}
A fraction of the irradiated $^{40}$Ar gas was transferred into stainless steel containers and shipped to the University of Bern. The gas volume (6.75 cm$^3$) was determined and subsequently diluted by a factor of 120 using bottled argon extracted from the atmosphere in 2018. 6\% of Methane was added as a quenching gas \citep{RIE16}. The $^{39}$Ar activity was measured in a 100 cc proportional counter operated at a pressure of 6500 mb. The LLC laboratory is located 35 meters underground, reducing the muon flux by a factor of 10. The $\beta$ energy deposited from $^{39}$Ar decays is recorded using a 7-bit Multi-Channel Analyzer (MCA). The detector background is determined through separate runs with argon depleted in $^{39}$Ar by a factor of 20--50, sourced from underground \citep{XUJ15}. Three measurements are performed to assess the $^{39}$Ar content of the sample: one using the depleted argon to quantify background, one with atmospheric argon as a reference, and one with the sample itself. The $^{39}$Ar activity of the sample is obtained by subtracting the measured background from the total activity and then comparing to the background-subtracted activity of atmospheric argon. Dilution corrected results (Table \ref{table-AMS}) are reported as the $^{39}$Ar/Ar ratio relative to the atmospheric reference value of $8.12\times 10^{-16}$.
\section{Results}
\label{results}
\begin{figure}[H]
	\centering 
	\includegraphics[width=0.7\textwidth, angle=0]{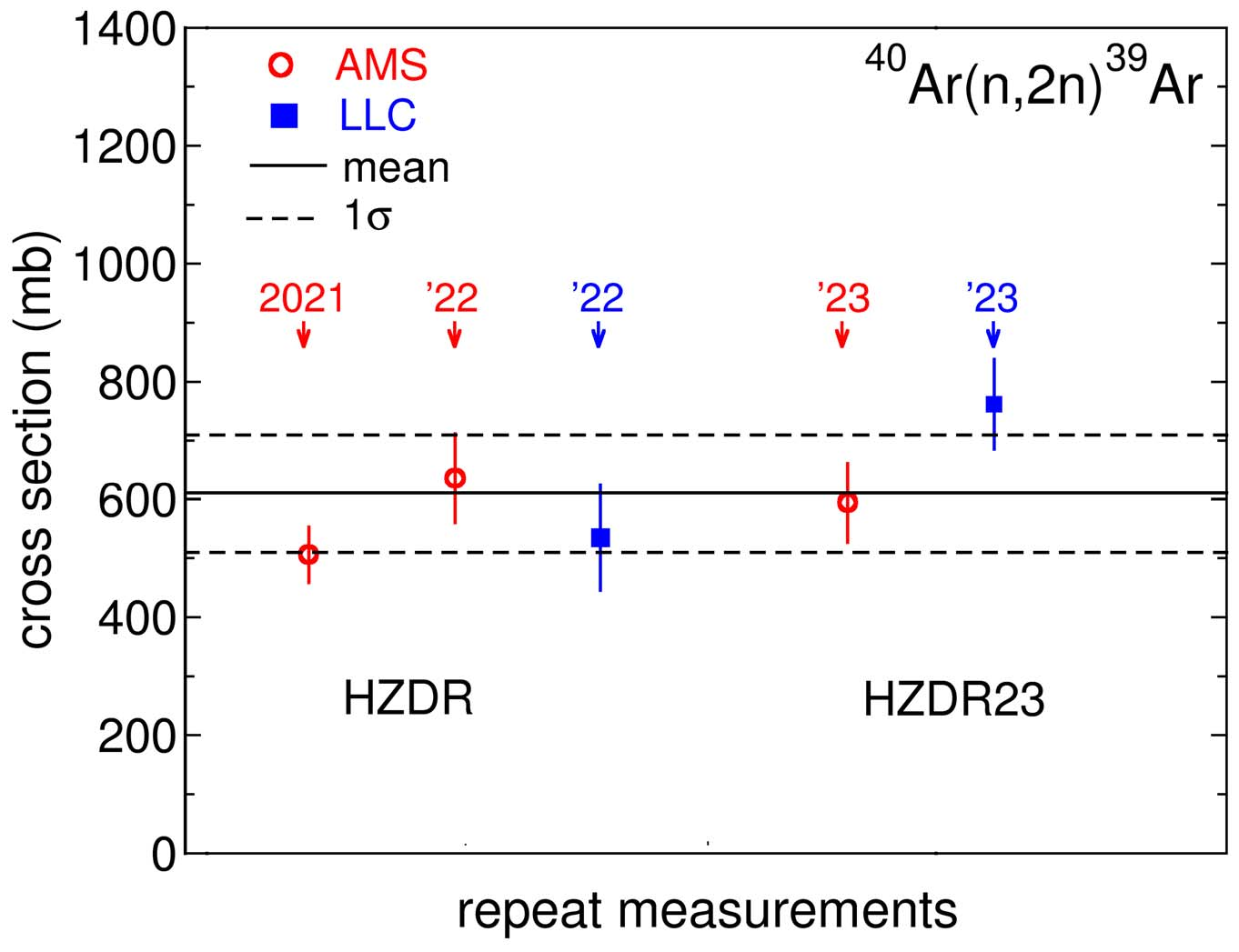}	
	\caption{Cross section of the $^{40}$Ar$(n,2n)^{39}$Ar reaction measured by AMS (open symbols) and LLC (solid symbols). The error bars represent 1$\sigma$ uncertainty derived from repeat measurements in each experiment. The adopted unweighted mean and standard deviation (dashed lines) of the measured cross section values are shown.} 
	\label{HZDR_result}%
\end{figure}
The cross section $\sigma_{39}$ of the $^{40}$Ar$(n,2n)^{39}$Ar reaction is extracted from the measured $^{39}$Ar/$^{40}$Ar isotopic ratios of Table \ref{table-AMS} using the expression $^{39}$Ar/$^{40}$Ar$= \Phi\cdot t\cdot\sigma_{39}$, where $\Phi\cdot t$ denotes the 14.8 MeV neutron fluence (Table \ref{irradiation}). The contribution of the $^{40}$Ar$(n,pn)^{39}$Cl reaction to the feeding of $^{39}$Ar, the cross section of which  was  measured to be 1.7(2) mb \citep{HUS1968}, is negligible. The unweighted mean and standard deviation of all our measured values of the $^{40}$Ar$(n,2n)^{39}$Ar reaction cross section (Fig. \ref{HZDR_result} and Table \ref{table-AMS}) are  610$\pm$100 mb at 14.8$\pm$0.3 MeV. Table \ref{uncertainty} lists the overall uncertainties for the sample HZDR23.
\begin{table}[H]
\centering
\caption{\label{uncertainty} Table of uncertainties (one sigma)} for HZDR23 measurements.
\vspace{0.3cm}
\begin{tabular}{lc}
\hline
\hline
source of uncertainty & Uncertainty ($\%$) \\
\hline
Ave. FC $^{40}$Ar$^{8+}$ & 3.8 \\
Stable beam attenuation & 3.0 \\
Ave. $^{39}$Ar rate-net & 4.7 \\
$^{39}$Ar fractionation correction & 4.9 \\
$^{39}$Ar/$^{40}$Ar -  \footnotesize{NOGAMS} & 8.3 \\
neutron fluence & 7.2 \\
cross sect. -  \footnotesize{NOGAMS} & 11.0 \\
\hline
count rate-net & 2.2 \\
atm-net & 10.3 \\
R/Ratm-counting gas & 9.8 \\
R/Ratm-sample & 8.2 \\
$^{39}$Ar/$^{40}$Ar -  \footnotesize{LLC} & 6.8 \\
cross sect. -  \footnotesize{LLC} & 10.0 \\
\hline
atmospheric neutron flux spectrum & 24$^a$ \\
lat., long., solar act., & 20$^b$ \\
overall uncertainty & 31\\
\hline
\end{tabular}
\begin{tablenotes}
        \item \footnotesize{$^a$ maximum uncertainty quoted for measured neutron spectrum  \citep{GOL00}}
        \item \footnotesize{$^b$ \citep{SLA11}}
  \end{tablenotes}
\end{table}

\begin{figure}[H] 
	\centering 
	\includegraphics[width=0.6\textwidth, angle=0]{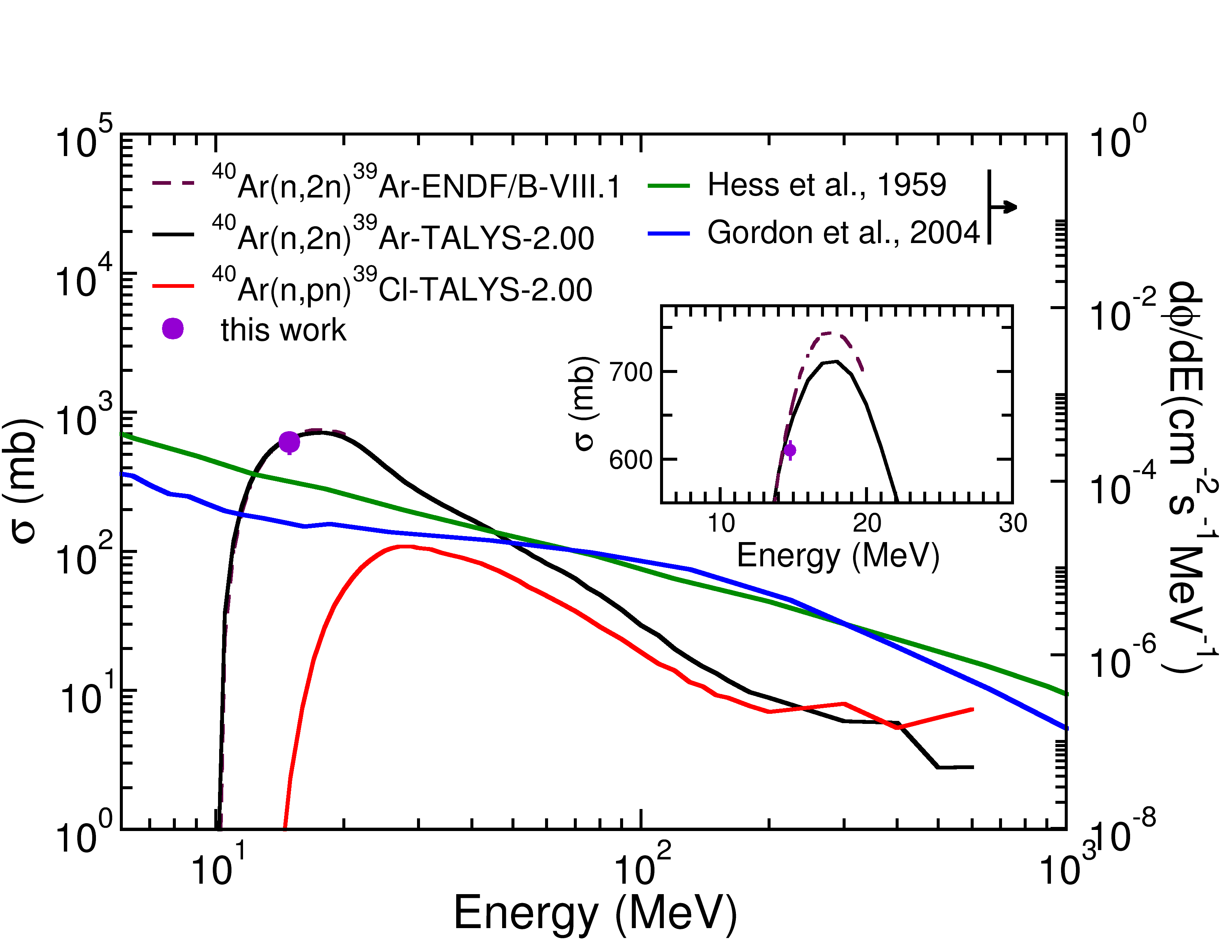}	
	\caption{Cross section (left axis) of the $^{40}$Ar$(n,2n)^{39}$Ar and $^{40}$Ar$(n,pn)^{39}$Cl reactions calculated by the ENDF/B-VIII.1 evaluation and the TALYS-2.00 code as a function of neutron energy. Our experimental value and the 1$\sigma$ uncertainty for the $^{40}$Ar$(n,2n)^{39}$Ar cross section at 14.8 MeV is plotted. The insert shows the region around the experimental point in linear scale. Experimental spectra (right axis) of neutron flux of cosmic origin at sea-level are also plotted.}
     
	\label{TENDL-TALYS}%
\end{figure}

\begin{table}[H]
\centering
\begin{threeparttable}
\caption{\label{production} $^{39}$Ar production rate $P$   at  different altitudes, calculated with Eq. \ref{eq1} } based on cosmic flux spectra from  \citep{GOL03, GOR04,HES59}.
\begin{tabular}{l c c}
\hline
\hline
altitude (km)&  $P$ ($10^3$ atoms/kg$_{Ar}$/day)& based on \\
\hline
0.2 & 0.744 & \citep{GOR04} \\
11.9 & 192 & \citep{GOL03}\\
16.2 & 474 & \\
20.0 & 530 & \\
\hline
0 & 1.47 & \citep{HES59} \\
3.2 & 112 & \\
12.2 & 237 & \\
\hline
\end{tabular}
\end{threeparttable}
\end{table}

\section{Discussion}
\label{discussion}
\subsection{Cosmogenic production of $^{39}$Ar in the atmosphere}
\begin{figure}[H]
	\centering 
	\includegraphics[width=0.5\textwidth, angle=0]{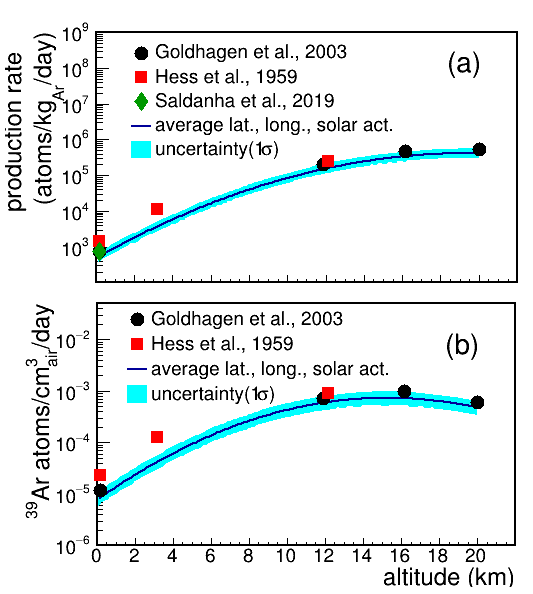}	
	\caption{(a) Cosmogenic atmospheric production rate of $^{39}$Ar at different altitudes calculated using energy-dependent cross sections of the TALYS-2.00 code (benchmarked by our experimental value of $^{40}$Ar$(n,2n)^{39}$Ar reaction cross section at 14.8 MeV, see Fig. \ref{TENDL-TALYS}) and spectra of cosmic-ray induced neutrons at different altitudes from   \citep{HES59,GOL03,GOR04} (solid symbols). The data are summed production rates via the $^{40}$Ar$(n,2n)^{39}$Ar and $^{40}$Ar$(n,pn)^{39}$Cl reactions. Our calculated value for sea-level production rate is in good agreement with the experimental value of \citep{SAL19} obtained with an artificial neutron flux spectrum.  The solid line represents a fit ($\bar{P}(z)$ in the text) of the production rates at the discrete locations and altitudes after averaging for dependence on latitude, longitude, and solar activity \citep{WOO19} (see text); the blue band denotes a 1$\sigma$ uncertainty partially based on \citep{GOL03}. (b) Production of $^{39}$Ar atoms per cm$^{3}$ of air per day at different altitudes (solid symbols), calculated using the production rates in (a). The solid line (blue band) is calculated using the averaged production rate ($1\sigma$ uncertainty) of (a).} 
	\label{Production-rate}%
\end{figure}

We note in Table \ref{cross-section} that values calculated by the recent version (2.00) the theoretical TALYS code \citep{KON23} and that of the ENDF/B-VIII.1 evaluation \citep{NOB24} are in good agreement with our experimental value of the 
$^{40}$Ar$(n,2n)^{39}$Ar reaction cross section; see also  
Fig. \ref{TENDL-TALYS} which shows the energy dependence of the theoretical TALYS-2.00 cross section \citep{KON23} of the $^{40}$Ar$(n,2n)^{39}$Ar and $^{40}$Ar$(n,pn)^{39}$Cl reactions. The production yield via the $(n,pn)$ reaction, negligible in our experiment at 14.8 MeV, amounts to $\approx$20\% of the $(n,2n)$ reaction over the whole energy range. The good agreement of the theoretical TALYS-2.00 value of the $^{40}$Ar$(n,2n)^{39}$Ar cross section with the experimental value at 14.8 MeV obtained in this work gives us confidence in using TALYS values to calculate the  atmospheric production rate of $^{39}$Ar at different altitudes from cosmogenic neutrons, using experimental neutron spectra \citep{HES59,GOL03,GOR04}; Fig. \ref{TENDL-TALYS} illustrates such spectra at sea level. The $^{39}$Ar production rate $P(z)$ (atoms/kg$_{Ar}$/day)  at  altitude $z$  from the interaction of cosmic ray neutrons can be written as,
\begin{equation}
\label{eq1}
P(z) = N_{Ar} \cdot n_{s/day} \int \frac{d\phi(E,z)}{dE} \sigma(E)\ dE,
\end{equation}
where $N_{Ar}$ is the number of Ar atoms per kg of argon, $n_{s/day}=86400$ the number of seconds in a day, $z$ the  altitude (cm), $d\phi(E,z)/dE$ is the cosmic-ray induced  neutron flux spectrum (neutrons/cm$^2$/s/MeV) and $\sigma(E)= \sigma_{(n,2n)}(E)+\sigma_{(n,pn)}(E)$ is the sum of the $^{40}$Ar$(n,2n)$ and $^{40}$Ar$(n,pn)$ cross sections (cm$^2$) calculated using the TALYS-2.00 code. Note that we use in this section the symbol Ar to denote either elemental Ar or $^{40}$Ar. The values of $P(z)$ are given in Table \ref{production} and shown in Fig. \ref{Production-rate}(a).  The production rates,  shown in Table \ref{production}, confirm the expected overwhelming contribution of $^{39}$Ar production at high altitudes compared to the sea-level value. In Fig. \ref{Production-rate}, we also average the 
$P(z)$ values extracted from the flux data of \citep{HES59,GOR04,GOL03} taken at different locations and times using the latitude, longitude and solar activity dependence of the neutron flux presented in \citep{WOO19}. This averaging, which we denote below by $\bar{P}(z)$, is justified by the short time scales (1--10 yr) of atmospheric exchanges \citep{HOB00} and solar activity (11 yr), compared to the $^{39}$Ar half-life (268 yr) and leads to consider the $^{39}$Ar abundance as homogeneous in the  atmosphere. 

We note  that the value determined here of 744 $^{39}$Ar atoms/kg$_{Ar}$/day for sea-level production rate is consistent with the experimental value ($759\pm128$  $^{39}$Ar atoms/kg$_{Ar}$/day) of \citep{SAL19}  who used an artificial neutron spectrum and scaled their values using different theoretical models. We show also in Fig. \ref{Production-rate}(b) the $^{39}$Ar production rates per cm$^3$ of air at different altitudes where the increased rates shown in Fig. \ref{Production-rate}(a) are mitigated at high altitudes by the lower air density.

Using the averaged altitude-dependent $^{39}$Ar production rate $\bar{P}(z)$ (atoms/kg$_{Ar}$/day), we can express
the total $^{39}$Ar production rate  $\bar{P}_{39Ar}$ in an air column of with a cross-sectional area of 1 cm$^2$ cross section as 
\begin{equation}{\label{P39ar}}
    \frac{d\bar{P}_{^{39}Ar}}{dS} = \int_{sea \ level}^{H_{max}} \bar{P}(z) \ m_{f} \ \rho_{air}(z) \ dz,
\end{equation}
where $m_f$ is the mass fraction of argon in air, and $\rho_{air}(z)$ is the air density (in kg/cm$^{3}$) at altitude $z$(cm), calculated from \citep{USS76}. The total  production rate, calculated from sea level up to an altitude of 20 km (denoted as $H_{max}$), is found to be $770\pm240$  $^{39}$Ar atoms/cm$^2$/day.
We can now calculate the  $^{39}$Ar content of the whole air column and its $^{39}$Ar/Ar isotopic ratio at secular equilibrium between cosmogenic production and radioactive decay. 
Implicit in this calculation is the negligible effect of chemical or physical sinks via chemical binding, adsorption, or dissolution \citep{OZI04}.
The steady-state  $^{39}$Ar/Ar ratio is expressed as:
\begin{equation}{\label{equ39Ar/Ar}}
     ^{39}Ar/Ar \ =\frac{(d\bar{P}_{^{39}Ar}/dS)/\lambda)}{dn_{Ar}/dS},
 \end{equation}
where $\lambda$ is the decay constant of $^{39}$Ar in day$^{-1}$ and $dn_{Ar}/dS$ denotes the number of Ar atoms per cm$^2$ of air column, calculated using the following relation:
\begin{equation}{\label{nAr}}
     dn_{Ar}/dS =\int_{sea \ level}^{H_{max}} \frac{ N_A \ m_{f} \ \rho_{air}(z)} {M_{Ar}} \ dz.
 \end{equation}
Here, $N_A$ denotes Avogadro's constant and $M_{Ar}$ is the molar mass of argon (in kg). 
Based on the experimental neutron flux spectra 
$d\phi(E,z)/dE$ from Goldhagen \textit{et al.} \citep{GOL03}, and using the $^{39}$Ar production rate calculated with Eq.\ref{P39ar}, the steady state  $^{39}$Ar/Ar ratio  from neutron-induced production in the atmosphere up to an altitude of 20 km is 
($5.9\pm1.8) \times 10^{-16}$. The contribution of the 
$^{40}$Ar$(n,pn)^{39}$Cl is $\approx$20\% of this value. Comparison with the experimental value of $^{39}$Ar/Ar= ($8.12\pm0.30)\times 10^{-16}$ in the atmosphere \citep{LOO83,BEN07shrt,GOL23} indicates that 73\% of $^{39}$Ar  is produced by cosmic-ray induced neutrons. This fraction is consistent with the estimate (72.3\%) of Saldanha \textit{et al.}  \citep{SAL19} based on sea-level production only. We note in this context that $^{39}$Ar production in shallow soils, dominated by the $^{39}$K$(n,p)^{39}$Ar reaction with cosmogenic neutrons, does not contribute significantly to the atmospheric $^{39}$Ar/Ar ratio \citep{FRI21,MUS23}. Muon and high-energy $\gamma$ rays induced reactions on $^{40}$Ar are considered in \citep{SAL19} to account for the major part of the remaining natural $^{39}$Ar production in the atmosphere. 

\subsection{Anthropogenic production of $^{39}$Ar in the atmosphere}
The presence of an anthropogenic contribution to atmospheric $^{39}$Ar, attributed to DT-neutrons from thermonuclear tests, was first noted by \citep{LOO68}. They compared the $^{39}$Ar specific activity in argon samples collected in 1940 and in 1959–1967, and estimated a contribution of $\lesssim$ 7\%. More recently, \citep{GU21} reconstructed the atmospheric $^{39}$Ar/Ar history and estimated this contribution to be $\lesssim$ 15\% of the total atmospheric $^{39}$Ar inventory.
The $^{40}$Ar$(n,2n)^{39}$Ar reaction cross section measured in this work at 14.8 MeV and its fair agreement with recent theoretical models (see Table \ref{cross-section} and Section \ref{discussion}) are essential data for a reliable estimate of the anthropogenic contribution to atmospheric $^{39}$Ar, originating mainly from the DT component of the thermonuclear devices. We calculate the $^{39}$Ar production via the $^{40}$Ar$(n,2n)^{39}$Ar reaction by simulating the transport of 14.1 MeV neutrons in the atmosphere using the GEANT4 code \citep{AGO03shrt} and the TALYS-2.00 code benchmarked in Section \ref{discussion}; a standard atmosphere composition is taken from \citep{USS76}. We find that a fraction of $1.5-3.4\times10^{-3}$ of emitted DT neutrons (14.1 MeV) produces $^{39}$Ar via the $(n,2n)$ reaction for high- to low-altitude tests, respectively. It is remarkable that the value of 2$\times10^{-3}$ originally quoted in \citep{LOO68} is fully consistent with this calculation. 
Using the neutron production of $8\times 10^{28}$ estimated by \citep{GU21} for all nuclear tests, our simulation yields between $1.2\times10^{26}$ and $2.7\times10^{26}$ anthropogenic $^{39}$Ar atoms, corresponding to $\approx$ 20\% of the atmospheric cosmogenic $^{39}$Ar inventory that is in qualitative agreement with the findings in \citep{GU21}.

\section{Conclusions}
Using 14.8 MeV neutrons from a DT neutron generator, we have determined for the first time the total cross section of the $^{40}$Ar$(n,2n)^{39}$Ar($t_{1/2}=$ 268 yr)  reaction as 610$\pm$100 mb. The cross section was determined by $^{39}$Ar atom counting with noble gas accelerator mass spectrometry and independently by low-level $\beta$ counting relative to atmospheric argon.  The recent TALYS-2.00  theoretical cross section values of the reaction  \citep{KON23} and the ENDF/B-VIII.1 evaluation \citep{NOB24}  are found to be in good agreement with our experimental value. We used the TALYS-2.00 energy-dependent theoretical cross section of the $^{40}$Ar$(n,2n)^{39}$Ar reaction, benchmarked at 14.8 MeV by our experimental value and that of the $^{40}$Ar$(n,pn)^{39}$Cl($t_{1/2}=56$ min) reaction (feeding long-lived $^{39}$Ar), together with experimental spectra of cosmic-ray induced neutrons, to calculate a production rate of 770 ± 240 $^{39}$Ar atoms/cm$^2$/day in the atmosphere  up to an altitude of 20 km. Comparison of the derived neutron-produced $^{39}$Ar/Ar isotopic abundance at secular equilibrium between production and decay with the experimental $^{39}$Ar/Ar atmospheric abundance shows that 73\% of atmospheric $^{39}$Ar originate in cosmic-ray induced neutrons, in agreement with an earlier result \citep{SAL19} based on sea-level production alone. This work establishes on firmer quantitative grounds the neutron-induced origin of the cosmogenic $^{39}$Ar abundance in the Earth's atmosphere. We use the experimental value of the $^{40}$Ar$(n,2n)^{39}$Ar reaction cross section for 14 MeV DT neutrons also to estimate the contribution of the thermonuclear tests of the 1960's to $^{39}$Ar in the atmosphere; this contribution is derived to be $\approx20$\% of the $^{39}$Ar atmospheric inventory.
\section{Data Availability}
Data are available through Zenodo at \\ 
https://doi.org/10.5281/zenodo.15974057 
\section*{Acknowledgements}
We thank the ATLAS Operation staff at Argonne National Laboratory for their dedication to this experiment. We thank R\"udiger Schanda for his help with the LLC measurements in Bern.  We gratefully acknowledge the support of USA-Israel Binational Science Foundation (BSF) under grant BSF-2020136, Pazy Foundation (Israel) and   EU program ChETEC-INFRA, 101008324, and BMFTR (05A23OD1).  This material is based upon work supported by the U.S. Department of Energy, Office of Science, Office of Nuclear Physics, under Contract No. DE-AC02-06CH11357. This research used resources of Argonne National Laboratory’s ATLAS facility, which is a DOE Office of Science User Facility.
This work is supported in part by National Science Foundation Grant No. NSF PHY-
2310059.
\end{linenumbers}
%
%
%

\bibliographystyle{elsarticle-harv}

\begin{thebibliography}{48}
\expandafter\ifx\csname natexlab\endcsname\relax\def\natexlab#1{#1}\fi
\providecommand{\url}[1]{\texttt{#1}}
\providecommand{\href}[2]{#2}
\providecommand{\path}[1]{#1}
\providecommand{\DOIprefix}{doi:}
\providecommand{\ArXivprefix}{arXiv:}
\providecommand{\URLprefix}{URL: }
\providecommand{\Pubmedprefix}{pmid:}
\providecommand{\doi}[1]{\href{http://dx.doi.org/#1}{\path{#1}}}
\providecommand{\Pubmed}[1]{\href{pmid:#1}{\path{#1}}}
\providecommand{\bibinfo}[2]{#2}
\ifx\xfnm\relax \def\xfnm[#1]{\unskip,\space#1}\fi
\bibitem[{Agostinelli et~al.(2003)Agostinelli, Allison, Amako, Apostolakis, Araujo et~al.}]{AGO03shrt}
\bibinfo{author}{Agostinelli, S.}, \bibinfo{author}{Allison, J.}, \bibinfo{author}{Amako, K.}, \bibinfo{author}{Apostolakis, J.}, \bibinfo{author}{Araujo, H.}, et~al., \bibinfo{year}{2003}.
\newblock \bibinfo{title}{{{GEANT4}} - a simulation toolkit}.
\newblock \bibinfo{journal}{Nucl. Instrum. Methods Phys. Res. Sect. A} \bibinfo{volume}{506}, \bibinfo{pages}{250 -- 303}.
\bibitem[{Alvarado et~al.(2007)Alvarado, Purtschert, Barbecot, Chabault, Rueedi, Schneider, Aeschbach‐Hertig, Kipfer and Loosli}]{COR07}
\bibinfo{author}{Alvarado, J.A.C.}, \bibinfo{author}{Purtschert, R.}, \bibinfo{author}{Barbecot, F.}, \bibinfo{author}{Chabault, C.}, \bibinfo{author}{Rueedi, J.}, \bibinfo{author}{Schneider, V.}, \bibinfo{author}{Aeschbach‐Hertig, W.}, \bibinfo{author}{Kipfer, R.}, \bibinfo{author}{Loosli, H.H.}, \bibinfo{year}{2007}.
\newblock \bibinfo{title}{Constraining the age distribution of highly mixed groundwater using {$^{39}$Ar}: A multiple environmental tracer {($^3$H/$^3$He, $^{85}$Kr, $^{39}$Ar, and $^{14}$C)} study in the semiconfined fontainebleau sands aquifer (france)}.
\newblock \bibinfo{journal}{Water Resour. Res.} \bibinfo{volume}{43}.
\bibitem[{Back and Ramachers(2008)}]{BAC08}
\bibinfo{author}{Back, J.}, \bibinfo{author}{Ramachers, Y.A.}, \bibinfo{year}{2008}.
\newblock \bibinfo{title}{{ACTIVIA}: Calculation of isotope production cross-sections and yields}.
\newblock \bibinfo{journal}{Nuclear Instruments and Methods in Physics Research Section A: Accelerators, Spectrometers, Detectors and Associated Equipment} \bibinfo{volume}{586}, \bibinfo{pages}{286--294}.
\bibitem[{Benetti et~al.(2007)Benetti, Calaprice, Calligarich, Cambiaghi, Carbonara et~al.}]{BEN07shrt}
\bibinfo{author}{Benetti, P.}, \bibinfo{author}{Calaprice, F.}, \bibinfo{author}{Calligarich, E.}, \bibinfo{author}{Cambiaghi, M.}, \bibinfo{author}{Carbonara, F.}, et~al., \bibinfo{year}{2007}.
\newblock \bibinfo{title}{Measurement of the specific activity of {$^{39}$Ar} in natural argon}.
\newblock \bibinfo{journal}{Nuclear Instruments and Methods in Physics Research Section A: Accelerators, Spectrometers, Detectors and Associated Equipment} \bibinfo{volume}{574}, \bibinfo{pages}{83--88}.
\bibitem[{Buizert et~al.(2014)Buizert, Baggenstos, Jiang, Purtschert, Petrenko, Lu, Müller, Kuhl, Lee, Severinghaus and Brook}]{BUI14}
\bibinfo{author}{Buizert, C.}, \bibinfo{author}{Baggenstos, D.}, \bibinfo{author}{Jiang, W.}, \bibinfo{author}{Purtschert, R.}, \bibinfo{author}{Petrenko, V.V.}, \bibinfo{author}{Lu, Z.T.}, \bibinfo{author}{Müller, P.}, \bibinfo{author}{Kuhl, T.}, \bibinfo{author}{Lee, J.}, \bibinfo{author}{Severinghaus, J.P.}, \bibinfo{author}{Brook, E.J.}, \bibinfo{year}{2014}.
\newblock \bibinfo{title}{{Radiometric $^{81}$Kr dating identifies 120,000-year-old ice at Taylor Glacier, Antarctica}}.
\newblock \bibinfo{journal}{Proceedings of the National Academy of Sciences} \bibinfo{volume}{111}, \bibinfo{pages}{6876--6881}.
\bibitem[{Callahan et~al.(2022)Callahan, Collon, Paul, Avila, Back et~al.}]{CAL22shrt}
\bibinfo{author}{Callahan, L.K.}, \bibinfo{author}{Collon, P.}, \bibinfo{author}{Paul, M.}, \bibinfo{author}{Avila, M.}, \bibinfo{author}{Back, B.}, et~al., \bibinfo{year}{2022}.
\newblock \bibinfo{title}{{Initial tests of Accelerator Mass Spectrometry with the Argonne Gas-Filled Analyzer and the commissioning of the MONICA detector}}.
\newblock \bibinfo{journal}{Nuclear Instruments and Methods in Physics Research Section B: Beam Interactions with Materials and Atoms} \bibinfo{volume}{532}, \bibinfo{pages}{7--12}.
\bibitem[{Cennini et~al.(1995)Cennini, Cittolin, {Dzialo Giudice}, Revol, Rubbia et~al.}]{CEN95shrt}
\bibinfo{author}{Cennini, P.}, \bibinfo{author}{Cittolin, S.}, \bibinfo{author}{{Dzialo Giudice}, D.}, \bibinfo{author}{Revol, J.}, \bibinfo{author}{Rubbia, C.}, et~al., \bibinfo{year}{1995}.
\newblock \bibinfo{title}{On atmospheric $^{39}${Ar} and $^{42}${Ar} abundance}.
\newblock \bibinfo{journal}{Nuclear Instruments and Methods in Physics Research Section A: Accelerators, Spectrometers, Detectors and Associated Equipment} \bibinfo{volume}{356}, \bibinfo{pages}{526--529}.
\bibitem[{Cerjan et~al.(2018)}]{CER18}
\bibinfo{author}{Cerjan, C.J.}, et~al., \bibinfo{year}{2018}.
\newblock \bibinfo{title}{{Dynamic high energy density plasma environments at the National Ignition Facility for nuclear science research}}.
\newblock \bibinfo{journal}{J. Phys. G: Nucl. Part. Phys.} \bibinfo{volume}{45}, \bibinfo{pages}{033003}.
\bibitem[{Chadwick et~al.(2011)Chadwick, Herman, Obložinský, Dunn, Danon, Kahler, Smith, Pritychenko, Arbanas, Arcilla et~al.}]{CHA11}
\bibinfo{author}{Chadwick, M.}, \bibinfo{author}{Herman, M.}, \bibinfo{author}{Obložinský, P.}, \bibinfo{author}{Dunn, M.}, \bibinfo{author}{Danon, Y.}, \bibinfo{author}{Kahler, A.}, \bibinfo{author}{Smith, D.}, \bibinfo{author}{Pritychenko, B.}, \bibinfo{author}{Arbanas, G.}, \bibinfo{author}{Arcilla, R.}, et~al., \bibinfo{year}{2011}.
\newblock \bibinfo{title}{{ENDF/B-VII}.1 nuclear data for science and technology: Cross sections, covariances, fission product yields and decay data}.
\newblock \bibinfo{journal}{Nuclear Data Sheets} \bibinfo{volume}{112}, \bibinfo{pages}{2887--2996}.
\bibitem[{Collon et~al.(2004)Collon, Bichler, Caggiano, Cecil, Masri, Golser, Jiang, Heinz, Henderson, Kutschera, Lehmann, Leleux, Loosli, Pardo, Paul, Rehm, Schlosser, Scott, {Smethie, Jr.} and Vondrasek}]{COL04a}
\bibinfo{author}{Collon, P.}, \bibinfo{author}{Bichler, M.}, \bibinfo{author}{Caggiano, J.}, \bibinfo{author}{Cecil, L.D.}, \bibinfo{author}{Masri, Y.E.}, \bibinfo{author}{Golser, R.}, \bibinfo{author}{Jiang, C.}, \bibinfo{author}{Heinz, A.}, \bibinfo{author}{Henderson, D.}, \bibinfo{author}{Kutschera, W.}, \bibinfo{author}{Lehmann, B.}, \bibinfo{author}{Leleux, P.}, \bibinfo{author}{Loosli, H.}, \bibinfo{author}{Pardo, R.}, \bibinfo{author}{Paul, M.}, \bibinfo{author}{Rehm, K.}, \bibinfo{author}{Schlosser, P.}, \bibinfo{author}{Scott, R.}, \bibinfo{author}{{Smethie, Jr.}, W.}, \bibinfo{author}{Vondrasek, R.}, \bibinfo{year}{2004}.
\newblock \bibinfo{title}{Development of an {AMS} method to study oceanic circulation characteristics using cosmogenic {$^{39}$Ar}}.
\newblock \bibinfo{journal}{Nucl. Instrum. Methods Phys. Res. B} \bibinfo{volume}{223-224}, \bibinfo{pages}{428 -- 434}.
\bibitem[{Davidenko et~al.(1957)Davidenko, Pogrebov and Saukov}]{DAV57GCA}
\bibinfo{author}{Davidenko, V.A.}, \bibinfo{author}{Pogrebov, I.S.}, \bibinfo{author}{Saukov, A.I.}, \bibinfo{year}{1957}.
\newblock \bibinfo{title}{Determination of the excitation function for the reaction {T(d,n)He$^4$}}.
\newblock \bibinfo{journal}{The Soviet Journal of Atomic Energy} \bibinfo{volume}{2}, \bibinfo{pages}{474--476}.
\bibitem[{Fritz et~al.(2021)Fritz, Alexander, Johnson, Mace, Milbrath and Hayes}]{FRI21}
\bibinfo{author}{Fritz, B.}, \bibinfo{author}{Alexander, T.}, \bibinfo{author}{Johnson, C.}, \bibinfo{author}{Mace, E.}, \bibinfo{author}{Milbrath, B.}, \bibinfo{author}{Hayes, J.}, \bibinfo{year}{2021}.
\newblock \bibinfo{title}{Background concentrations of argon-39 in shallow soil gas}.
\newblock \bibinfo{journal}{Journal of Environmental Radioactivity} \bibinfo{volume}{228}, \bibinfo{pages}{106513}.
\bibitem[{Goldhagen(2000)}]{GOL00}
\bibinfo{author}{Goldhagen, P.}, \bibinfo{year}{2000}.
\newblock \bibinfo{title}{Overview of aircraft radiation exposure and recent {ER-2} measurements}.
\newblock \bibinfo{journal}{Health Physics} \bibinfo{volume}{79}, \bibinfo{pages}{526--544}.
\bibitem[{Goldhagen et~al.(2003)Goldhagen, Clem and Wilson}]{GOL03}
\bibinfo{author}{Goldhagen, P.}, \bibinfo{author}{Clem, J.}, \bibinfo{author}{Wilson, J.}, \bibinfo{year}{2003}.
\newblock \bibinfo{title}{Recent results from measurements of the energy spectrum of cosmic-ray induced neutrons aboard an {ER-2} airplane and on the ground}.
\newblock \bibinfo{journal}{Advances in Space Research} \bibinfo{volume}{32}, \bibinfo{pages}{35--40}.
\bibitem[{Golovko(2023)}]{GOL23}
\bibinfo{author}{Golovko, V.V.}, \bibinfo{year}{2023}.
\newblock \bibinfo{title}{Application of the most frequent value method for {$^{39}$Ar} half-life determination}.
\newblock \bibinfo{journal}{Eur. Phys. J. C} \bibinfo{volume}{83}, \bibinfo{pages}{930}.
\bibitem[{Gordon et~al.(2004)Gordon, Goldhagen, Rodbell, Zabel, Tang, Clem and Bailey}]{GOR04}
\bibinfo{author}{Gordon, M.}, \bibinfo{author}{Goldhagen, P.}, \bibinfo{author}{Rodbell, K.}, \bibinfo{author}{Zabel, T.}, \bibinfo{author}{Tang, H.}, \bibinfo{author}{Clem, J.}, \bibinfo{author}{Bailey, P.}, \bibinfo{year}{2004}.
\newblock \bibinfo{title}{Measurement of the flux and energy spectrum of cosmic-ray induced neutrons on the ground}.
\newblock \bibinfo{journal}{IEEE Transactions on Nuclear Science} \bibinfo{volume}{51}, \bibinfo{pages}{3427--3434}.
\bibitem[{Gu et~al.(2021)Gu, Tong, Yang, Hu, Jiang, Lu, Purtschert and Ritterbusch}]{GU21}
\bibinfo{author}{Gu, J.Q.}, \bibinfo{author}{Tong, A.L.}, \bibinfo{author}{Yang, G.M.}, \bibinfo{author}{Hu, S.M.}, \bibinfo{author}{Jiang, W.}, \bibinfo{author}{Lu, Z.T.}, \bibinfo{author}{Purtschert, R.}, \bibinfo{author}{Ritterbusch, F.}, \bibinfo{year}{2021}.
\newblock \bibinfo{title}{Reconstruction of the atmospheric {$^{39}$Ar/Ar} history}.
\newblock \bibinfo{journal}{Chemical Geology} \bibinfo{volume}{583}, \bibinfo{pages}{120480}.
\bibitem[{Hess et~al.(1959)Hess, Patterson, Wallace and Chupp}]{HES59}
\bibinfo{author}{Hess, W.N.}, \bibinfo{author}{Patterson, H.W.}, \bibinfo{author}{Wallace, R.}, \bibinfo{author}{Chupp, E.L.}, \bibinfo{year}{1959}.
\newblock \bibinfo{title}{Cosmic-ray neutron energy spectrum}.
\newblock \bibinfo{journal}{Phys. Rev.} \bibinfo{volume}{116}, \bibinfo{pages}{445--457}.
\bibitem[{Hobbs(2000)}]{HOB00}
\bibinfo{author}{Hobbs, P.V.}, \bibinfo{year}{2000}.
\newblock \bibinfo{title}{Introduction to atmospheric chemistry}.
\newblock \bibinfo{publisher}{Cambridge University Press}.
\bibitem[{Hou et~al.(2025)Hou, Jenk, Jiang, Zhang, Hu, Feng, Li, Wu, Pang, Yu, Huang, Lu, Yang, Bender and Schwikowski}]{HOU25}
\bibinfo{author}{Hou, S.}, \bibinfo{author}{Jenk, T.M.}, \bibinfo{author}{Jiang, W.}, \bibinfo{author}{Zhang, W.}, \bibinfo{author}{Hu, H.}, \bibinfo{author}{Feng, X.}, \bibinfo{author}{Li, H.}, \bibinfo{author}{Wu, S.Y.}, \bibinfo{author}{Pang, H.}, \bibinfo{author}{Yu, J.}, \bibinfo{author}{Huang, R.}, \bibinfo{author}{Lu, Z.T.}, \bibinfo{author}{Yang, G.M.}, \bibinfo{author}{Bender, M.}, \bibinfo{author}{Schwikowski, M.}, \bibinfo{year}{2025}.
\newblock \bibinfo{title}{A radiometric timescale challenges the chronology of the iconic 1992 {Guliya} ice core}.
\newblock \bibinfo{journal}{Science Advances} \bibinfo{volume}{11}, \bibinfo{pages}{eadx8837}.
\bibitem[{Husain and Kuroda(1968)}]{HUS1968}
\bibinfo{author}{Husain, L.}, \bibinfo{author}{Kuroda, P.}, \bibinfo{year}{1968}.
\newblock \bibinfo{title}{14·8 {MeV} neutron activation cross-sections of argon}.
\newblock \bibinfo{journal}{Journal of Inorganic and Nuclear Chemistry} \bibinfo{volume}{30}, \bibinfo{pages}{355--359}.
\bibitem[{Iwamoto et~al.(2023)Iwamoto, Iwamoto, Kunieda, Minato, Nakayama, Abe, Tsubakihara, Okumura, Ishizuka, Yoshida et~al.}]{IWA23}
\bibinfo{author}{Iwamoto, O.}, \bibinfo{author}{Iwamoto, N.}, \bibinfo{author}{Kunieda, S.}, \bibinfo{author}{Minato, F.}, \bibinfo{author}{Nakayama, S.}, \bibinfo{author}{Abe, Y.}, \bibinfo{author}{Tsubakihara, K.}, \bibinfo{author}{Okumura, S.}, \bibinfo{author}{Ishizuka, C.}, \bibinfo{author}{Yoshida, T.}, et~al., \bibinfo{year}{2023}.
\newblock \bibinfo{title}{Japanese evaluated nuclear data library version 5: {JENDL}-5}.
\newblock \bibinfo{journal}{Journal of Nuclear Science and Technology} \bibinfo{volume}{60}, \bibinfo{pages}{1--60}.
\bibitem[{Klix et~al.(2018)Klix, D\"oring, Domula and Zuber}]{KLI18}
\bibinfo{author}{Klix, A.}, \bibinfo{author}{D\"oring, T.}, \bibinfo{author}{Domula, A.}, \bibinfo{author}{Zuber, K.}, \bibinfo{year}{2018}.
\newblock \bibinfo{title}{The intensive {DT} neutron generator of {TU Dresden}}.
\newblock \bibinfo{journal}{EPJ Web of Conferences} \bibinfo{volume}{170}, \bibinfo{pages}{02004}.
\bibitem[{Koning et~al.(2023)Koning, Hilaire and Goriely}]{KON23}
\bibinfo{author}{Koning, A.J.}, \bibinfo{author}{Hilaire, S.}, \bibinfo{author}{Goriely, S.}, \bibinfo{year}{2023}.
\newblock \bibinfo{title}{{TALYS}: modeling of nuclear reactions}.
\newblock \bibinfo{journal}{Eur. Phys. J. A} \bibinfo{volume}{59}, \bibinfo{pages}{131}.
\bibitem[{Kostal et~al.(2023)Kostal, Czakoj, Alexa, {\v Simon}, {Zme\v skal}, Schulc, {Krechlerov\'a}, {Pelt\'an}, Mravec, Cvachovec, Rypar, {Uhl\'ar}, Harkut and {Mat\v ej}}]{KOS23}
\bibinfo{author}{Kostal, M.}, \bibinfo{author}{Czakoj, T.}, \bibinfo{author}{Alexa, P.}, \bibinfo{author}{{\v Simon}, J.}, \bibinfo{author}{{Zme\v skal}, M.}, \bibinfo{author}{Schulc, M.}, \bibinfo{author}{{Krechlerov\'a}, A.}, \bibinfo{author}{{Pelt\'an}, T.}, \bibinfo{author}{Mravec, F.}, \bibinfo{author}{Cvachovec, F.}, \bibinfo{author}{Rypar, V.}, \bibinfo{author}{{Uhl\'ar}, R.}, \bibinfo{author}{Harkut, O.}, \bibinfo{author}{{Mat\v ej}, Z.}, \bibinfo{year}{2023}.
\newblock \bibinfo{title}{Measurement of dosimetrical cross sections with 14.05 {MeV} neutrons from compact neutron generator}.
\newblock \bibinfo{journal}{Annals of Nuclear Energy} \bibinfo{volume}{191}, \bibinfo{pages}{109904}.
\bibitem[{Kutschera et~al.(2023)Kutschera, Jull, Paul and Wallner}]{KUT23}
\bibinfo{author}{Kutschera, W.}, \bibinfo{author}{Jull, A.J.T.}, \bibinfo{author}{Paul, M.}, \bibinfo{author}{Wallner, A.}, \bibinfo{year}{2023}.
\newblock \bibinfo{title}{Atom counting with accelerator mass spectrometry}.
\newblock \bibinfo{journal}{Rev. Mod. Phys.} \bibinfo{volume}{95}, \bibinfo{pages}{035006}.
\bibitem[{Loosli(1983)}]{LOO83}
\bibinfo{author}{Loosli, H.}, \bibinfo{year}{1983}.
\newblock \bibinfo{title}{A dating method with {$^{39}$Ar}}.
\newblock \bibinfo{journal}{Earth Planet. Sci. Lett.} \bibinfo{volume}{63}, \bibinfo{pages}{51 -- 62}.
\bibitem[{Loosli and Oeschger(1968)}]{LOO68}
\bibinfo{author}{Loosli, H.}, \bibinfo{author}{Oeschger, H.}, \bibinfo{year}{1968}.
\newblock \bibinfo{title}{Detection of {$^{39}$Ar} in atmospheric argon}.
\newblock \bibinfo{journal}{Earth and Planetary Science Letters} \bibinfo{volume}{5}, \bibinfo{pages}{191--198}.
\bibitem[{MacMullin et~al.(2012)MacMullin, Boswell, Devlin, Elliott, Fotiades, Guiseppe, Henning, Kawano, B.H.LaRoque, Nelson and O'Donnell}]{MAC12}
\bibinfo{author}{MacMullin, S.}, \bibinfo{author}{Boswell, M.}, \bibinfo{author}{Devlin, M.}, \bibinfo{author}{Elliott, S.}, \bibinfo{author}{Fotiades, N.}, \bibinfo{author}{Guiseppe, V.}, \bibinfo{author}{Henning, R.}, \bibinfo{author}{Kawano, T.}, \bibinfo{author}{B.H.LaRoque}, \bibinfo{author}{Nelson, R.}, \bibinfo{author}{O'Donnell, J.}, \bibinfo{year}{2012}.
\newblock \bibinfo{title}{{Partial $\ensuremath{\gamma}$-ray production cross sections for ($n,\phantom{\rule{-0.16em}{0ex}}xn\ensuremath{\gamma}$) reactions in natural argon at 1--30 MeV}}.
\newblock \bibinfo{journal}{Phys. Rev. C} \bibinfo{volume}{85}, \bibinfo{pages}{064614}.
\bibitem[{Mancusi et~al.(2014)Mancusi, Boudard, Cugnon, David, Kaitaniemi and Leray}]{MAN14}
\bibinfo{author}{Mancusi, D.}, \bibinfo{author}{Boudard, A.}, \bibinfo{author}{Cugnon, J.}, \bibinfo{author}{David, J.C.}, \bibinfo{author}{Kaitaniemi, P.}, \bibinfo{author}{Leray, S.}, \bibinfo{year}{2014}.
\newblock \bibinfo{title}{Extension of the {Li{\`e}ge} intranuclear-cascade model to reactions induced by light nuclei}.
\newblock \bibinfo{journal}{Physical Review C} \bibinfo{volume}{90}, \bibinfo{pages}{054602}.
\bibitem[{Musy and Purtschert(2023)}]{MUS23}
\bibinfo{author}{Musy, S.}, \bibinfo{author}{Purtschert, R.}, \bibinfo{year}{2023}.
\newblock \bibinfo{title}{Reviewing {$^{39}$Ar} and {$^{37}$Ar} underground production in shallow depths with implications for groundwater dating}.
\newblock \bibinfo{journal}{Science of The Total Environment} \bibinfo{volume}{884}, \bibinfo{pages}{163868}.
\bibitem[{Nassar et~al.(2005)Nassar, Paul, Ahmad, Berkovits, Bettan et~al.}]{NAS05shrt}
\bibinfo{author}{Nassar, H.}, \bibinfo{author}{Paul, M.}, \bibinfo{author}{Ahmad, I.}, \bibinfo{author}{Berkovits, D.}, \bibinfo{author}{Bettan}, et~al., \bibinfo{year}{2005}.
\newblock \bibinfo{title}{Stellar $(n,\ensuremath{\gamma})$ cross section of $^{62}\mathrm{N}\mathrm{i}$}.
\newblock \bibinfo{journal}{Phys. Rev. Lett.} \bibinfo{volume}{94}, \bibinfo{pages}{092504}.
\bibitem[{Nobre et~al.(2024)Nobre, Brown, Arcilla, Coles and Shu}]{NOB24}
\bibinfo{author}{Nobre, G.}, \bibinfo{author}{Brown, D.}, \bibinfo{author}{Arcilla, R.}, \bibinfo{author}{Coles, R.}, \bibinfo{author}{Shu, B.}, \bibinfo{year}{2024}.
\newblock \bibinfo{title}{Progress towards the {ENDF/B-VIII.1} release}.
\newblock \bibinfo{journal}{EPJ Web of Conf.} \bibinfo{volume}{294}, \bibinfo{pages}{04004}.
\bibitem[{Ozima and Podosek(2004)}]{OZI04}
\bibinfo{author}{Ozima, M.}, \bibinfo{author}{Podosek, F.A.}, \bibinfo{year}{2004}.
\newblock \bibinfo{title}{Noble Gas Geochemistry (Second Edition)}.
\newblock \bibinfo{publisher}{Cambridge University Press}.
\bibitem[{Paul et~al.(1989)Paul, Glagola, Henning, Keller, Kutschera, Liu, Rehm, Schneck and Siemssen}]{PAU89}
\bibinfo{author}{Paul, M.}, \bibinfo{author}{Glagola, B.G.}, \bibinfo{author}{Henning, W.}, \bibinfo{author}{Keller, J.G.}, \bibinfo{author}{Kutschera, W.}, \bibinfo{author}{Liu, Z.}, \bibinfo{author}{Rehm, K.E.}, \bibinfo{author}{Schneck, B.}, \bibinfo{author}{Siemssen, R.H.}, \bibinfo{year}{1989}.
\newblock \bibinfo{title}{Heavy ion separation with a gas-filled magnetic spectrograph}.
\newblock \bibinfo{journal}{Nucl. Instrum. Methods Phys. Res. A} \bibinfo{volume}{277}, \bibinfo{pages}{418 -- 430}.
\bibitem[{Paul et~al.(2019)Paul, Pardo, Collon, Kutschera, Rehm, Scott and Vondrasek}]{PAU19}
\bibinfo{author}{Paul, M.}, \bibinfo{author}{Pardo, R.C.}, \bibinfo{author}{Collon, P.}, \bibinfo{author}{Kutschera, W.}, \bibinfo{author}{Rehm, K.E.}, \bibinfo{author}{Scott, R.}, \bibinfo{author}{Vondrasek, R.C.}, \bibinfo{year}{2019}.
\newblock \bibinfo{title}{Positive-ion accelerator mass spectrometry at {ATLAS}: Peaks and pits}.
\newblock \bibinfo{journal}{Nucl. Instrum. Methods Phys. Res. Sec. B} \bibinfo{volume}{456}, \bibinfo{pages}{222 -- 229}.
\bibitem[{Riedmann and Purtschert(2016)}]{RIE16}
\bibinfo{author}{Riedmann, R.A.}, \bibinfo{author}{Purtschert, R.}, \bibinfo{year}{2016}.
\newblock \bibinfo{title}{Separation of argon from environmental samples for {Ar-37} and {Ar-39} analyses}.
\newblock \bibinfo{journal}{Separation and Purification Technology} \bibinfo{volume}{170}, \bibinfo{pages}{217--223}.
\bibitem[{Saldanha et~al.(2019)Saldanha, Back, Tsang, Alexander, Elliott, Ferrara, Mace, Overman and Zalavadia}]{SAL19}
\bibinfo{author}{Saldanha, R.}, \bibinfo{author}{Back, H.O.}, \bibinfo{author}{Tsang, R.H.M.}, \bibinfo{author}{Alexander, T.}, \bibinfo{author}{Elliott, S.R.}, \bibinfo{author}{Ferrara, S.}, \bibinfo{author}{Mace, E.}, \bibinfo{author}{Overman, C.}, \bibinfo{author}{Zalavadia, M.}, \bibinfo{year}{2019}.
\newblock \bibinfo{title}{Cosmogenic production of $^{39}\mathrm{Ar}$ and $^{37}\mathrm{Ar}$ in argon}.
\newblock \bibinfo{journal}{Phys. Rev. C} \bibinfo{volume}{100}, \bibinfo{pages}{024608}.
\bibitem[{{Silberberg} and {Tsao}(1973a)}]{SIL73I}
\bibinfo{author}{{Silberberg}, R.}, \bibinfo{author}{{Tsao}, C.H.}, \bibinfo{year}{1973}a.
\newblock \bibinfo{title}{{Partial Cross-Sections in High-Energy Nuclear Reactions, and Astrophysical Applications. I. Targets With {Z}$\leq$ 28.}}
\newblock \bibinfo{journal}{Astrophysical Journal Supplement} \bibinfo{volume}{25}, \bibinfo{pages}{315--333}.
\bibitem[{{Silberberg} and {Tsao}(1973b)}]{SIL73II}
\bibinfo{author}{{Silberberg}, R.}, \bibinfo{author}{{Tsao}, C.H.}, \bibinfo{year}{1973}b.
\newblock \bibinfo{title}{{Partial Cross-Sections in High-Energy Nuclear Reactions, and Astrophysical Applications. II. Targets Heavier than Nickel}}.
\newblock \bibinfo{journal}{Astrophysical Journal Supplement} \bibinfo{volume}{25}, \bibinfo{pages}{335--367}.
\bibitem[{Slayman(2011)}]{SLA11}
\bibinfo{author}{Slayman, C.}, \bibinfo{year}{2011}.
\newblock \bibinfo{title}{JEDEC Standards on Measurement and Reporting of Alpha Particle and Terrestrial Cosmic Ray Induced Soft Errors}. \bibinfo{publisher}{Springer, US}, \bibinfo{address}{Boston, MA}.
\newblock pp. \bibinfo{pages}{55--76}.
\bibitem[{Stoenner et~al.(1965)Stoenner, Schaeffer and Katcoff}]{STO65}
\bibinfo{author}{Stoenner, R.W.}, \bibinfo{author}{Schaeffer, O.A.}, \bibinfo{author}{Katcoff, S.}, \bibinfo{year}{1965}.
\newblock \bibinfo{title}{{Half-Lives of Argon-37, Argon-39, and Argon-42}}.
\newblock \bibinfo{journal}{Science} \bibinfo{volume}{148}, \bibinfo{pages}{1325}.
\bibitem[{Tessler et~al.(2018)Tessler, Paul, Halfon, Meyer, Pardo, Purtschert et~al.}]{TES18shrt}
\bibinfo{author}{Tessler, M.}, \bibinfo{author}{Paul, M.}, \bibinfo{author}{Halfon, S.}, \bibinfo{author}{Meyer, B.S.}, \bibinfo{author}{Pardo, R.}, \bibinfo{author}{Purtschert}, et~al., \bibinfo{year}{2018}.
\newblock \bibinfo{title}{Stellar $^{36,38}\mathrm{Ar}(n,\ensuremath{\gamma})^{37,39}\mathrm{Ar}$ reactions and their effect on light neutron-rich nuclide synthesis}.
\newblock \bibinfo{journal}{Phys. Rev. Lett.} \bibinfo{volume}{121}, \bibinfo{pages}{112701}.
\bibitem[{{United States Committee on Extension to the Standard Atmosphere}(1976)}]{USS76}
\bibinfo{author}{{United States Committee on Extension to the Standard Atmosphere}}, \bibinfo{year}{1976}.
\newblock \bibinfo{title}{U.S. Standard Atmosphere, 1976}.
\newblock NOAA-S/T 76-1562, \bibinfo{publisher}{National Oceanic and Atmospheric Administration}, \bibinfo{address}{Washington, D.C.}
\bibitem[{W.~Mannhart(2007)}]{MAN07}
\bibinfo{author}{W.~Mannhart, D.S.}, \bibinfo{year}{2007}.
\newblock \bibinfo{title}{Measurement of neutron activation cross sections in the energy range from 8 {MeV} to 15 {MeV}}.
\newblock \bibinfo{journal}{{PTB-Bericht Bremerhaven, Wirtschaftsverl. NW, Verl. für neue Wiss.,}} \bibinfo{volume}{{N 53}}.
\bibitem[{Woolf et~al.(2019)Woolf, Sinclair, {Van Brabant}, Harvey, Phlips, Hutcheson and Jackson}]{WOO19}
\bibinfo{author}{Woolf, R.S.}, \bibinfo{author}{Sinclair, L.E.}, \bibinfo{author}{{Van Brabant}, R.A.}, \bibinfo{author}{Harvey, B.J.}, \bibinfo{author}{Phlips, B.F.}, \bibinfo{author}{Hutcheson, A.L.}, \bibinfo{author}{Jackson, E.G.}, \bibinfo{year}{2019}.
\newblock \bibinfo{title}{Measurement of secondary cosmic-ray neutrons near the geomagnetic north pole}.
\newblock \bibinfo{journal}{Journal of Environmental Radioactivity} \bibinfo{volume}{198}, \bibinfo{pages}{189--199}.
\bibitem[{Xu et~al.(2015)Xu, Calaprice, Galbiati, Goretti, Guray, Hohman, Holtz, Ianni, Laubenstein, Loer, Love, Martoff, Montanari, Mukhopadhyay, Nelson, Rountree, Vogelaar and Wright}]{XUJ15}
\bibinfo{author}{Xu, J.}, \bibinfo{author}{Calaprice, F.}, \bibinfo{author}{Galbiati, C.}, \bibinfo{author}{Goretti, A.}, \bibinfo{author}{Guray, G.}, \bibinfo{author}{Hohman, T.}, \bibinfo{author}{Holtz, D.}, \bibinfo{author}{Ianni, A.}, \bibinfo{author}{Laubenstein, M.}, \bibinfo{author}{Loer, B.}, \bibinfo{author}{Love, C.}, \bibinfo{author}{Martoff, C.}, \bibinfo{author}{Montanari, D.}, \bibinfo{author}{Mukhopadhyay, S.}, \bibinfo{author}{Nelson, A.}, \bibinfo{author}{Rountree, S.}, \bibinfo{author}{Vogelaar, R.}, \bibinfo{author}{Wright, A.}, \bibinfo{year}{2015}.
\newblock \bibinfo{title}{A study of the trace {$^{39}$Ar} content in argon from deep underground sources}.
\newblock \bibinfo{journal}{Astroparticle Physics} \bibinfo{volume}{66}, \bibinfo{pages}{53--60}.
\bibitem[{Yokochi et~al.(2012)Yokochi, Sturchio and Purtschert}]{YOK12}
\bibinfo{author}{Yokochi, R.}, \bibinfo{author}{Sturchio, N.C.}, \bibinfo{author}{Purtschert, R.}, \bibinfo{year}{2012}.
\newblock \bibinfo{title}{Determination of crustal fluid residence times using nucleogenic {$^{39}$Ar}}.
\newblock \bibinfo{journal}{Geochimica et Cosmochimica Acta} \bibinfo{volume}{88}, \bibinfo{pages}{19--26}.

\end{thebibliography}

\end{document}